\documentclass[aps,prl,twocolumn,groupedaddress,showpacs]{revtex4-1}
\usepackage{graphicx}
\begin{document}
\title{Transient behavior of full counting statistics in thermal transport}

\author{Jian-Sheng Wang}
\author{Bijay Kumar Agarwalla}
\author{Huanan Li}
\affiliation{Department of Physics and Centre for Computational Science and Engineering, National University of Singapore, Singapore 117542, Republic of Singapore} 

\date{24 June 2011}

\begin{abstract}
The generating function of energy counting statistics is derived
for phononic junction systems.  It is expressed in terms of the
contour-ordered self-energy of the lead with shifted arguments,
$\Sigma^A(\tau, \tau') = \Sigma_L\bigl(\tau +\hbar x(\tau), \tau' +
\hbar x(\tau')\bigr) - \Sigma_L(\tau, \tau')$, where $\Sigma_L(\tau,\tau')$
is the usual contour-ordered self-energy of the left lead.  The
cumulants of the energy transferred in a given time $t_M$ from the
lead to the center is obtained by taking derivatives.  A transient
result of the first four cumulants of a graphene junction is
presented.  It is found that measurements cause the energy to flow
into the lead.
\end{abstract}


\maketitle

Phonon transport in the ballistic quantum regime possesses special
features, such as the quantized universal thermal conductance
\cite{Rego-Kirczenow-1998,Schwab-2000} and wave-like coherent
transport described by a Landauer-like formula
\cite{Yamamoto-2006,WangJS-europhysJb-2008}.  A typical set-up of such
a system consists of two infinite heat baths maintained at
different temperatures with a finite junction part forming the
scattering region.  The focus in the last decade has been on 
steady-state thermal currents.  Since the heat baths are stochastic in
nature, it is natural to ask a statistical question: 
what is the
distribution of the energy $Q$ transferred in a given time $t_M$.
Such questions have been raised in electron transport, where it is
known as the full counting statistics. Levitov and Lesovik presented
their celebrated formula which forms the definite answer to the
question \cite{Levitov}. Many works followed in electronic transport
\cite{otherworks,Esposito-review-2009}.  The electron counting
statistics has been experimentally measured in quantum dot systems
\cite{Flindt-etal-2009}.  No such measurements have been carried out
for thermal transport, but it is potentially possible, e.g., in a
nano-resonator system.

Saito and Dhar \cite{Saito-Dhar-2007} treated the full counting
statistics for heat transport in a 1D chain. Such inquiries also have
deep connections with the nonequilibrium fluctuation theorems
\cite{noneq-fluct}.  The result obtained by Saito and Dhar was only
for the long-time limit.  In this paper, we present a formulation
based on two-time measurements, treating the transient behavior and
long-time limit on an equal footing.  A central result of our
derivation is that the generating function can be concisely expressed
by the contour-ordered self-energies of the lead, making contact with
the nonequilibrium Green's function (NEGF) method
\cite{WangJS-europhysJb-2008} of quantum transport.  A more general
expression for the long-time limit of a general junction system with
any number of degrees of freedom is also derived, and numerical
results for the transient behavior of the first few cumulants of a
graphene junction are presented.
 
We consider initially decoupled harmonic systems described by the
Hamiltonians
\begin{equation}
H_\alpha = \frac{1}{2} p_\alpha^T p_\alpha + \frac{1}{2} u_\alpha^T K^\alpha u_\alpha, \quad \alpha=L, R, C,
\end{equation}
for the left and right leads and a central region.  The leads are
assumed semi-infinite while the center has a finite number of degrees
of freedom.  Masses are absorbed by defining $u = \sqrt{m}\,
x$. $u_\alpha$ and $p_\alpha$ are column vectors of coordinates and
momenta. $K^\alpha$ is the spring constant matrix of region $\alpha$.
Couplings of the center region with the leads are turned on either
adiabatically from time $t=-\infty$, or switched on abruptly at $t=0$.
The interaction term takes the form $H_{{\rm int}} = u_L^T V^{LC}u_{C}
+ u_R^T V^{RC} u_{C}$.  The total Hamiltonian is $H = H_L + H_C + H_R
+ H_{\rm int}$.

Focusing on the left lead, we define the energy current operator by
the rate of decrease of energy of the lead (in the Heisenberg picture)
as
\begin{equation}
I(t) = - \frac{dH_L(t)}{dt}=\frac{i}{\hbar}[H_L, H_H]= p_L(t)^TV^{LC}u_C(t),
\end{equation}
where $H_H$ is the Hamiltonian in the Heisenberg picture.  We define
the `heat' operator as
\begin{equation}
\label{eq-hatQ}
\hat Q=\int_0^t\! I(t')\,dt' = H_L - U(0,t) H_L U(t,0),
\end{equation}
where $H_L$ [$=H_L(0)$] is the Schr\"odinger operator of the free left
lead, and $U(t,t')$ is the evolution operator under $H(t)$.  $U$
satisfies the Schr\"odinger equation
\begin{equation}
i\hbar { \partial U(t,t') \over \partial t} = H(t) U(t,t').
\end{equation}

What we would like to calculate is the moments of the heat energy
transferred in a given time $t$.  To this end, we look at the
generating function of the moments instead.  Since $\hat Q$ is a
quantum operator, there are subtleties as to how exactly this
generating function should be defined.  Na\"ively, we may use $\langle
e^{i\xi \hat{Q}} \rangle$.  But this definition fails the fundamental
requirement of positive definiteness of the probability distribution,
\begin{equation}
P(Q) = \int e^{-i \xi Q} Z(\xi) \frac{d\xi}{2\pi},
\end{equation}
for a classical quantity
$Q$.  The correct definition is \cite{Esposito-review-2009,talkner2008}
\begin{equation}
Z = \langle e^{i\xi H_L} e^{-i\xi H_L(t)}\rangle',
\end{equation}
based on measurements at time 0 and $t$ where each time a measurement
of the energy of the left lead is carried out, the wavefunction collapses
into the eigenstate of the operator $H_L$.  Thus, to take care of this
process, the average is defined by
\begin{equation}
\langle \cdots \rangle' = {\rm Tr} \Bigl[ \sum_{a} P_a\, \rho(0) P_a  \cdots \Bigr],
\end{equation}
where $P_a$ is the projector onto the eigenstate of $H_L$ with
eigenvalue $a$. $\rho(0)$ is the steady-state density operator
obtained by adiabatically evolving from a product state at $t=-\infty$
to $t=0$.

To calculate the generating function $Z$, we use the following
strategies.  First, the projector is represented by Fourier transform,
$P_a = \delta(a-H_L) = \int_{-\infty}^\infty e^{-i\lambda(a-H_L)}
d\lambda/(2\pi)$.  Then, the products of the exponential factors in
$Z$, combined with the exponential factors in the projectors, are
written in terms of an evolution operator $U_x(t,t')$ of an effective
Hamiltonian with a parameter $x$, given
\begin{eqnarray}
 Z(\xi) &=& \langle e^{i \xi H_L/2} e^{-i \xi H_L(t) } e^{i \xi H_L/2 }  \rangle' \nonumber \\
&\propto& \int \frac{d\lambda}{2\pi} {\rm Tr} \bigl\{ 
\rho(0) U_{\xi/2-\lambda}(0,t) U_{-\xi/2-\lambda}(t,0) \bigr\} 
\label{eq-Uxilambda} \nonumber \\
&= &  \int \frac{d\lambda}{2\pi}\; Z(\xi, \lambda).
\end{eqnarray}
The proportionality constant will be fixed later by the condition
$Z(0)=1$.  The evolution operator $U_{x}$ is associated with the
Hamiltonian
\begin{eqnarray}
H_{x}(t) &=& e^{ix H_L} H(t) e^{-i x H_L} \nonumber \\
& = & H(t) + \bigl( u_L(\hbar x) - u_L\bigr)^T V^{LC} u_C,
\end{eqnarray}
where $u_L(\hbar x) = e^{i x H_L} u_L e^{-i x H_L}$ is the free left lead
``Heisenberg'' evolution to time $t = \hbar x$.  We can give a more explicit form for the Hamiltonian,
\begin{equation}
H_x(t) = H(t) + \bigl[ u_L^T C(x) + p_L^T S(x) \bigr]u_C,
\end{equation}
where 
\begin{eqnarray}
\label{eq-C}
C(x) &=& \bigl(\cos(\hbar x \sqrt{K_L}) -1\bigr) V^{LC}, \\
\label{eq-S}
S(x) &=& (1/\sqrt{K_L}) \sin( \hbar x \sqrt{K_L}) V^{LC}.
\end{eqnarray}
Next, we represent $U_x$ using path integrals.  The lagrangians
associated with the path integrals are (ignoring the right lead for
the moment):
\begin{eqnarray}
{\cal L}_L &=& \frac{1}{2}\dot{u}_L^2 - \frac{1}{2} u_L^T K^L u_L, \\
{\cal L}_C &=& \frac{1}{2} \dot{u}_C^2 - \frac{1}{2} u_C^T \bigl(K^C-S^TS\bigr) u_C, \\
{\cal L}_{LC} &=& - \dot{u}^T_L S u_C - u_L^T \bigl(V^{LC} + C\bigr) u_C.
\end{eqnarray}
Following Feynman and Vernon \cite{Feynman-Vernon-1963}, we can
eliminate the leads by performing gaussian integrals.  Since the
coupling to the center is linear, the result will be a quadratic form
in the exponential, i.e., another gaussian.  The 
influence functional is given by
\begin{eqnarray}
I[u_C(\tau)] &\equiv & \int {\cal D}[u_L] \rho_L(-\infty) 
e^{\frac{i}{\hbar} \int d\tau ({\cal L}_L + {\cal L}_{LC}) } \nonumber  \\
&=& {\rm Tr}\Bigl[\frac{e^{-\beta_L H_L}}{Z_L} T_c e^{ -\frac{i}{\hbar} \int d\tau V_I(\tau) } \Bigr] \nonumber \\
&=& e^{-\frac{i}{2\hbar} \int\int d\tau d\tau' u_C^T(\tau) \Pi(\tau, \tau') u_C(\tau') },\\
V_I(\tau)  &=&u_L^T\bigl(\tau + \hbar x(\tau)\bigr)V^{LC}u_C + 
\frac{1}{2} u_C^T S^T S u_C.\quad
\end{eqnarray}  
In the above expressions, the contour function $u_C(\tau)$ is not a
dynamical variable but only a parametric function.  $T_c$ is the
contour order operator.  Note that $V_I$ is the interaction picture
operator with respect to $H_L$, as a result, $e^{itH_L/\hbar }
u_{L}(\hbar x) e^{-itH_L/\hbar} =u_L(t + \hbar x)$.  We define the
contour function $x(\tau)$ as 0 whenever $t < 0$ or $t>t_M$. 
Otherwise it is $x^{+}(t) = -\xi/2 - \lambda$ on the upper branch, and
$x^{-}(t) = \xi/2 - \lambda$ on the lower branch.  The important
influence functional self-energy on the contour is
\begin{eqnarray}
\Pi(\tau,\tau') &=& \Sigma_L^A + \Sigma_L + S^TS\delta(\tau, \tau'), \\
\Sigma^A + \Sigma_L &= & V^{CL} g_L\bigl(\tau + \hbar x(\tau), \tau'+\hbar x(\tau') \bigr) V^{LC} \nonumber \\
\label{eq-SAL}
&=&  \Sigma_L\bigl(\tau + \hbar x(\tau), \tau'+\hbar x(\tau') \bigr),
\end{eqnarray}
where $\Sigma_L$ is the usual lead contour self-energy, $\delta$ is
the Dirac delta function defined on the contour. Equation
(\ref{eq-SAL}) is the most important equation defining the self-energy
of the problem.  The generating function $Z$ can be expressed in terms
of the usual Green's function $G=G^0_{CC}$ of the central region and
this particular self-energy.  The self-energy $\Sigma^A$ is obtained
from the lead self-energy $\Sigma_L$ by appropriately shifting the
contour time arguments and taking a difference.  With this result,
infinite degrees of freedom (due to the semi-infinite nature of the
leads) reduce to finite degrees of freedom.

The generating function is obtained by another gaussian integral, given
\begin{eqnarray}
Z(\xi,\lambda)  
&=& \int {\cal D}[u_C] \rho_C(-\infty) e^{(i/\hbar) 
\int d\tau  {\cal L}_C } I[u_C] \nonumber \\
&=& \int {\cal D}[u_C] \rho_C(-\infty) e^{\frac{i}{\hbar}S_{\rm eff}} \nonumber \\
&\propto & {\rm det}(D)^{-1/2},
\end{eqnarray}
where
\begin{eqnarray}
S_{\rm eff} &=& \frac{1}{2} \int d\tau \int d\tau' u_C^T(\tau) D(\tau, \tau') u_C(\tau'),  \\
D(\tau, \tau') &=& - \frac{\partial^2}{\partial \tau^2} \delta(\tau, \tau')- K^C  \delta(\tau, \tau') \nonumber\\ 
&& \> -\Sigma(\tau, \tau') - \Sigma^A(\tau, \tau') \nonumber \\
&=& D_0 - \Sigma^A,
\end{eqnarray}
where $\Sigma = \Sigma_L + \Sigma_R$.
We define the Green's function $G$ by $D_0 G = 1$, or more precisely
\begin{equation}
\int D_0(\tau, \tau'') G(\tau'', \tau') d\tau'' = \delta(\tau, \tau').
\end{equation}
In the above formula for $Z$, we imagine that the differential
operator (integral operator) $D$ and $D_0^{-1}$ are represented as
matrices indexed by space $j$ and contour time $\tau$. We can make a
systematic expansion in term of $\Sigma^A$ by noting the following
formulae for matrices, ${\rm det}(M) = e^{{\rm Tr} \ln M}$, and $\ln
(1 - y) = -\sum_{k=1}^\infty \frac{y^k}{k}.$ Using this, we can write
\begin{equation}
\ln Z(\xi) = \lim_{\lambda \to \infty} \sum_{k=1}^\infty \frac{1}{2k}
{\rm Tr}_{(j,\tau)} \Bigl[(G \Sigma^A)^k \Bigr].
\label{Z-final-formula}
\end{equation}
This formula is the central result of this paper.  The expression is
valid for any transient time $t_M$ embedded in the self-energy
$\Sigma^A$.  The notation ${\rm Tr}_{(j,\tau)}$ means trace both in
space $j$ and contour time $\tau$, i.e., integrating over the Keldysh
contour.  The projection to the eigenstates of $H_L$ results in an
integration over $\lambda$.  Since the range of the integration is
from $-\infty$ to $+\infty$, and the two-parameter generating function
$Z(\xi, \lambda)$ approaches a constant as $| \lambda| \to \infty$,
the value of the integral is dominated by the value at infinity.  Our
choice of the proportionality factor satisfies the required condition
of $Z(0) = 1$.

For NEGF notations and relations among Green's functions, we refer to
Ref.~\cite{WangJS-europhysJb-2008}.  It is more convenient to work
with a Keldysh rotation for the contour ordered functions, keeping
${\rm Tr}(AB \cdots C)$ invariant.  For any $A^{\sigma\sigma'}(t,t')$,
with $\sigma, \sigma'=\pm$ for branch indices, the effect of the
Keldysh rotation is to change to
\begin{eqnarray}
\breve{A} &=& 
\left( \begin{array}{cc}
                             A^r & A^K \\
                             A^{\bar K} & A^a 
\end{array} \right)  \\
 &=&
\frac{1}{2} \left( \begin{array}{cc}
              A^t - A^{\bar t} - A^< + A^>, & A^t + A^{\bar t} + A^< + A^> \\
              A^t + A^{\bar t} - A^< - A^>, & A^t - A^{\bar t} + A^< - A^>  
                         \end{array} \right). \nonumber
\end{eqnarray}
We should view the above as defining the quantities 
$A^r$, $A^a$, $A^K$, and $A^{\bar{K}}$.   
For the usual Green's function $G$ we get
\begin{equation}
\breve{G} = 
\left( \begin{array}{cc}
                             G^r & G^K \\
                             0 & G^a 
\end{array} \right) .
\end{equation}
The $G^{\bar K}$ component is 0 due to the standard relation among the
Green's functions.  But the $\bar K$ component is nonzero for
$\Sigma^A$.

In the long-time limit, translational invariance is restored for the
self-energies.  Convolution in time domain simply becomes
multiplication in the frequency domain. The shifts given to the
arguments in $\Sigma_L$ become independent of time $t$, only depend on
the branches.  We have
\begin{eqnarray}
\Sigma_A^t &=& \Sigma_A^{\bar t} = 0,\\
\Sigma_A^<(t) &=& \Sigma_L^{<}(t-\hbar \xi) -\Sigma_L^{<}(t),\\
 \Sigma_A^>(t) &=& \Sigma_L^{>}(t+\hbar \xi)-\Sigma_L^{>}(t).
\end{eqnarray}
Fourier transforming the lesser and greater self-energies, we obtain
$\Sigma_A^{<}[\omega] = \Sigma^{<}_L[\omega] \bigl(e^{i\hbar \omega
\xi} - 1 \bigr)$, $\Sigma_A^{>}[\omega] = \Sigma^{>}_L[\omega]
\bigl(e^{-i\hbar \omega \xi} - 1 \bigr)$.  We can now compute the
matrix product $\breve{G} \breve{\Sigma}^A$. Finally, the generating
function for large $t_M$ is
\begin{eqnarray}
\ln Z(\xi) &= &  - t_M \int_{-\infty}^{+\infty}\!\! \frac{d\omega}{4\pi} {\rm Tr} \ln \bigl(1 - \breve{G} \breve{\Sigma}^A \bigr) \nonumber \\
&\!\!\!\!\!\!\!\!=&\!\!\!\!  -t_M \int_{-\infty}^{+\infty}\!\!\frac{d\omega}{4\pi} \ln \det \Bigl\{ 1 - G^r \Gamma_L 
G^a \Gamma_R \big[  (e^{i\xi \hbar \omega}\! -\! 1) f_L \nonumber \\ 
&&\!\!\!\!\!\!\!\!\!\!\!\!+ ( e^{-i\xi\hbar \omega} \!-\! 1) f_R + (
e^{i\xi \hbar \omega} \!+\! e^{-i\xi\hbar\omega} \!-\!2 ) f_L f_R \big]\Bigr\}.\qquad
\label{eq-lnZxi}
\end{eqnarray}
where $\breve{G}$, $\breve{\Sigma}^A$, and $\Gamma_\alpha =
i(\Sigma_\alpha^r -\Sigma_\alpha^a)$ are in the frequency domain and
$f_{\alpha} = 1/\bigl(e^{\beta_\alpha \hbar \omega } - 1\bigr)$,
$\beta_\alpha = 1/(k_B T_\alpha)$, is the Bose distribution function.
This result generalizes that of Saito and Dhar
\cite{Saito-Dhar-2007}. It satisfies the steady-state fluctuation
theorem \cite{fluct-theorems}, $Z(\xi) = Z\bigl(-\xi + i (\beta_R -
\beta_L)\bigr)$.

The long-time result does not depend on how the initial states are
prepared before measurement.  This is not the case for transience.  The
generating function, Eq.~(\ref{Z-final-formula}), is for the case
where the system is prepared in a steady state.  A measurement at time
0 disturbs the system, and similarly at time $t_M$.  Instead of a
steady state, we can also prepare the system in a product state,
$\rho(-\infty) \propto \exp(-\sum_\alpha \beta_\alpha H_\alpha)$.  This
means that the coupling $H_{\rm int}$ is switched on suddenly.  Then
the projector $P_a$ commutes with the density matrix with no effect on
$\rho(-\infty)$.  This simplifies the problem.  We use the Feynman
diagrammatic technique to obtain the result. Omitting the details, we
have
\begin{eqnarray}
\label{eq-Zprod}
\ln Z_0 &=& - \frac{1}{2} {\rm Tr}_{(j,\tau)} \ln \big( 1 - G_0 \Sigma^{A} \big).
\end{eqnarray}
This expression looks formally the same as before except that $G_0$
satisfies a Dyson equation defined on the contour from 0 to $t_M$ and
back, while $G$ is defined on the Keldysh contour from $-\infty$ to
$t_M$.
\begin{eqnarray}
G_0(\tau, \tau') &=& g_C(\tau,\tau') \\
\label{eq-Dyson}
&&\> + \int \!\int d\tau_1d \tau_2\, 
g_C(\tau, \tau_1) \Sigma(\tau_1,  \tau_2) G_0(\tau_2, \tau'), \nonumber
\end{eqnarray}
where $g_C$ is the contour ordered Green's function of the isolated center.

\begin{figure}
\includegraphics[width=\columnwidth]{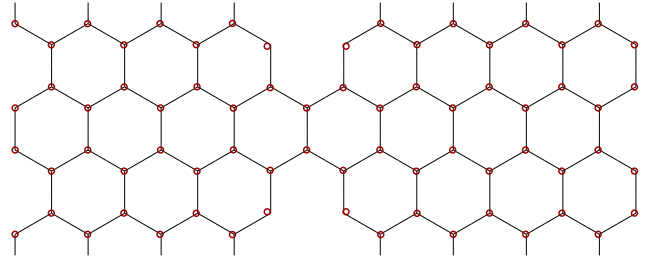}%
\caption{The structure of a graphene junction with 6 degrees of
freedom with two carbon atoms as the center.}
\end{figure}
\begin{figure}
\includegraphics[width=1.05\columnwidth]{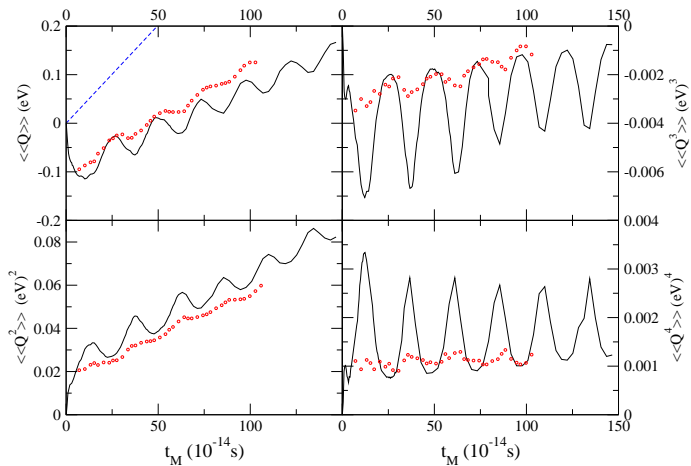}%
\caption{The cumulants $\langle \langle Q^n \rangle \rangle$ for
$n=1$, 2, 3 and 4.  The curves are for the product initial state; the
circles are for steady-state initial state.  The dotted line
is for the classical limit ($\hbar
\to 0$ keeping $\lambda$ finite) for the steady-state initial
condition.  The temperature of the left lead is 330 K and that of
the right lead is 270 K. For the product initial state, the center
temperature is 300 K.}
\end{figure}

We now present some numerical results.  Fig.~1 is the structure of our
graphene junction system.  The center region consists of two atoms,
while the two leads are symmetrically arranged as strips (with
periodic boundary conditions in the vertical direction).  We obtained
the force constants using the second generation Brenner potential.  To
compute the transient results, we need to perform convolution
integrations in the time or frequency domain many times.  It is handled by
treating the convolutions as matrix multiplications.  Then the
expression of the derivatives, $\langle\langle Q^n \rangle \rangle
=\partial^n \ln Z/\partial (i\xi)^n$, is calculated.  Note that the
$\xi$ dependence only enters through $\Sigma^A$.  We also note that a
power series in $\breve{G} \breve{\Sigma}^A$ terminates after $n$
terms for $\langle\langle Q^n \rangle \rangle$ for the product-state
initial condition, but it is an infinite series for the steady-state
case.  The computational effort required for convergence is huge for
the graphene junction.  We also obtained the result for 1D chain
which will be presented elsewhere.

Fig.~2 shows the first four cumulants.  The first cumulant, which is
also the first moment, is the total amount of energy entering the
center from the left lead during time 0 to $t_M$.  Its derivative
gives the current.  Such transient currents have been calculated
\cite{eduardos-paper} for the product initial states for 1D chains.
The second cumulant gives the variance of $Q$.  The higher order
cumulants are small but not zero, thus the distribution of $Q$ is not
gaussian. For large times, all the cumulants become linear
in $t_M$, and are in agreement to the long-time prediction.

One striking feature of the results is that the product initial state
and the steady-state initial state results behave
qualitatively the same.  The heat transferred, $\langle Q \rangle$,
starts from 0 and goes down to negative values.  This means whether we
start from a decoupled system or a steady state, the effect of
measurement is always to feed energy into the measured (left) lead,
even if the temperature of the left lead is lower than that of the
right lead.  If the system were classical, the measurement cannot
disturb the system. We should expect the current to be 
constant once the steady state is established.  The nonlinear $t_M$
dependence observed here in $\langle Q\rangle$ is fundamentally
quantum-mechanical in origin.  

In summary, the generating function for phononic junction systems is
obtained, which can be written compactly using Green's function as
$\ln Z = - (1/2) {\rm Tr} \ln (1 - G \Sigma^A)$.  A central quantity
is the self-energy $\Sigma^A$ which is expressed in terms of the usual lead
self-energy with shifted arguments.  This is a very general result
valid for steady-state initial states or product initial states in a
two-time measurement.  Numerical results for a graphene junction
system are presented.  An intriguing feature is that a measurement,
even in the steady state, causes energy to flow into the leads.  We
hope that such robust features can be verified experimentally.

This work is supported in part by a URC research grant
R-144-000-257-112 of National University of Singapore.

%



\end{document}